\documentclass[twocolumn,amssymb, nobibnotes, showpacs, superscriptaddress, aps, pra]{revtex4-1}
\usepackage{graphicx,amsmath}

\begin{document}


\title{Weak-Light, Zero to $-\pi$ Lossless Kerr-Phase Gate in Quantum-well System via Tunneling Interference Effect}


\author{Y. L. Shi}
\affiliation{School of Physics Science and Engineering, Tongji University, Shanghai 200092, China}

\author{Y. C. Huang}
\affiliation{School of Physics Science and Engineering, Tongji University, Shanghai 200092, China}

\author{J. X. Wu}
\affiliation{School of Physics Science and Engineering, Tongji University, Shanghai 200092, China}

\author{C. J. Zhu}
\email[]{cjzhu@tongji.edu.cn}
\affiliation{School of Physics Science and Engineering, Tongji University, Shanghai 200092, China}
\affiliation{MOE Key Laboratory of Advanced Micro-Structured Materials, School of Physics Science and Engineering, Tongji University, Shanghai 200092, China}

\author{J. P. Xu}
\affiliation{School of Physics Science and Engineering, Tongji University, Shanghai 200092, China}
\affiliation{MOE Key Laboratory of Advanced Micro-Structured Materials, School of Physics Science and Engineering, Tongji University, Shanghai 200092, China}

\author{Y. P. Yang}
\affiliation{School of Physics Science and Engineering, Tongji University, Shanghai 200092, China}
\affiliation{MOE Key Laboratory of Advanced Micro-Structured Materials, School of Physics Science and Engineering, Tongji University, Shanghai 200092, China}


\date{\today}

\begin{abstract}
We examine a Kerr phase gate in a semiconductor quantum well structure based on the tunnelling interference effect. We show that there exist a specific signal field detuning, at which the absorption/amplification of the probe field will be eliminated with the increase of the tunnelling interference. Simultaneously, the probe field will acquire a $-\pi$ phase shift at the exit of the medium. We demonstrate with numerical simulations that a complete $180^\circ$ phase rotation for the probe field at the exit of the medium is achieved, which may result in many applications in information science and telecommunication.
\end{abstract}

\pacs{42.65.Hw, 42.50.Gy, 78.67.De}

\maketitle
Highly efficient optical-field manipulation protocols are critically important to quantum computers which hold the promise to revolutionize the information science~\cite{Nielsen2000,Gisin2007,Gisin2002,Mermin2007,Vedral2007,Matthews2013,Jaeger2007}. The operation of controlling qubits with qubits is the key technique of the protocols. To this end, many proposals have come up in recent years for efficiently implementing all-optical quantum computation. One of the preferred and also widely discussed schemes to achieve such controlling qubits with qubits  is the phase gate at single photon level or a few photons level~\cite{Rauschenbeutel1999,Giovannetti2000,Englert2001,Schmidt2003,Gasparoni2004,Brien2003}.  There are many proposals~\cite{Crespi2011,Vitali2000,Ottaviani2003,Lukin2001,Resch2002}  based on various phenomena such as magnetic field, light-field induced shift, nonlinear effects such as Kerr cross-phase modulations, and so on. The phase gate based on Kerr cross-phase modulations, i.e. the Kerr phase gate, brings the real possibility of achieving a true manipulation of polarization-encoded light field at single or a few photon level in a confined hollow-core optical fiber, which has been demonstrated experimentally and discussed theoretically in atomic system~\cite{Li2013,Li2014,Zhu2014}.


As we all known, the Kerr nonlinearity is important not only for most nonlinear optical processes~\cite{Shen} but also for many applications in quantum information processing, including quantum nondemolition measurement, quantum state teleportation, quantum logic gates and others~\cite{nie}. Under normal circumstances the Kerr nonlinearity is produced in passive optical media such as glass-based optical fibers, in which far-off resonance excitation schemes are used to avoid optical absorption. As a result, the Kerr nonlinearity is too small to enhance the photon-photon interaction so the optical quantum phase gate operation cannot be efficiently implemented. With the advent of the electromagnetic induced transparency (EIT)~\cite{Fleischhauer}, Kerr nonlinearity can be greatly enhanced in the presence of the quantum interference if a system works near resonance in atomic systems. Up to now, many schemes based on EIT such as N scheme~\cite{Lukin2000}, four-level inverted-Y scheme~\cite{Xiao} and other variations have been used in theoretical studies of the enhancement of the Kerr nonlinearity, which are predicted to be good candidate to realize quantum entanglement of ultraslow photons~\cite{Lukin2000}, single photon switching~\cite{Harris1998}, nonlinear phase gate~\cite{Ottaviaani2003}, and single photon propagation controls~\cite{Lukin2001}.

However, EIT-based Kerr nonlinearity scheme has some drawbacks that are difficult to overcome. The primary question of the weakly driven EIT-based scheme is that the probe field undergoes a significant attenuation even if it works in the transparency window. Although the Kerr nonlinearity can be greatly enhanced when the system works near resonance, the third-order attenuation is also significantly boosted because of the ultraslow propagation~\cite{Deng2007}. For this reason, it was generally recognized that EIT-based schemes are unrealistic to take advantages of such resonantly enhanced Kerr nonlinearity. Several experimental studies based on EIT scheme have shown small nonlinear Kerr phase shifts using cold atomic gases~\cite{Kang2003}. To overcome these drawbacks, an active Raman gain (ARG) scheme was proposed to realize large and rapidly responding Kerr nonlinearity enhancement at room temperature~\cite{Deng2007}, which eliminates the significant probe field attenuation or distortion associated with weakly driven EIT-based schemes. Recently, a fast-responsed, Kerr phase gate and polarization gate have been demonstrated experimentally in the ARG-based atomic system\cite{Li2013,Li2014}.

\begin{figure}[htb]
  \centering
  \includegraphics[width=8cm]{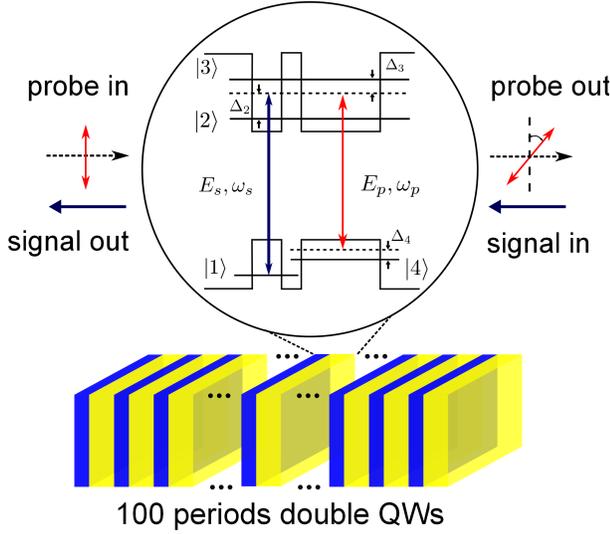}\\
  \caption{(Color online) The geometry of 100 periods of asymmetric double quantum wells and the configuration of the light field. Each asymmetric double quantum wells consists of a narrow well and a wide well separated by a small tunnelling barrier. $|1\rangle$ and $|4\rangle$ are localized hole states of valance band. $|2\rangle$ and $|3\rangle$ are delocalized bounding and anti-bounding states, which are coupled by a thin tunneling barrier. $E_{s(p)}$ is the electric field of the signal (probe) field with the angular frequency $\omega_{s(p)}$. $\Delta_j\ (j=2-4)$ are the detunings of state $|j\rangle$. }\label{fig:model}
\end{figure}
In this work, we examine a lossless Kerr phase shift via tunnelling interference effect in an asymmetric double quantum-well (QW) structure. The QW system (see Fig.~\ref{fig:model}) interacts with a weak signal field, which is used as a phase-control field, and a probe field at single/few photon level simultaneously. We show that the attenuation of the signal field can be greatly reduced because of the tunnelling interference effect, which is known as tunneling induced transparency (TIT). We point out that the physical mechanism of TIT is based on destructive interference. Prior to this theoretical study, it was widely believed that the TIT results from the Autler-Townes splitting. We further show that there exist a specific signal field detuning, at which the absorption/amplification of the probe field will be eliminated with the increase of the tunnelling interference. Simultaneously, the probe field will acquire a $-\pi$ Kerr phase shift at the exit of the medium. We demonstrate with numerical simulations that a complete $180^\circ$ phase rotation for the probe field at the exit of the medium is achieved when the signal field is turned on. However, the probe field propagates through the medium without any phase accumulation when the signal field is turned off.


%

As show in Fig.~\ref{fig:model}, we consider a multiple quantum-well structure which consists of $100$ periods of GaAs/Al$_{0.5}$Ga$_{0.5}$As double quantum well. Each double quantum well starts with (from left to right)~\cite{QWref,Zhu2009,Zhu2011} a thick Al$_\textrm{0.5}$Ga$_{0.5}$As barrier layer that is followed by a GaAs layer of thickness of $25$ monolayers (ML). This narrow well is separated from a $60$ ML GaAs layer (a wide well) by a $9$ ML Al$_\textrm{0.5}$Ga$_{0.5}$As potential barrier layer. Finally, a thick Al$_\textrm{0.5}$Ga$_{0.5}$As barrier layer is added on the right of the wide well to separate it from other double quantum wells. Each ML in the region between the narrow and wide wells is made of GaAs material with two-dimensional electronic density ${\cal{N}}_e=3\times10^{11}\ {\rm cm}^{-2}$~\cite{Zhu2009,Zhu2011} and has a thickness of 0.28 nm (see Fig.~\ref{fig:model}). In this structure, the first electron state in the conduction band of the wide well is energetically aligned with that of the narrow well, whereas the first hole states in the valence bands of both wells are not aligned with each other. Because of the small tunneling barrier, electrons delocalize and the corresponding states split into a bounding and an anti-bounding states (labeled as $|2\rangle$ and $|3\rangle$ respectively). The holes remain localized and the corresponding states are labeled as $|1\rangle$ and $|4\rangle$, respectively.

To achieve a zero to $-\pi$ lossless Kerr-phase gate operation at weak light level, a weak signal field $E_s$ with angular frequency $\omega_s$ couples the fully occupied ground state $|1\rangle$ and the upper states $|2\rangle$ and $|3\rangle$ simultaneously. The half-Rabi frequencies of the signal field are defined as $\Omega_{21}=(\mathbf{p}_{21}\cdot\mathbf{E}_s)/(2\hbar)$ and $\Omega_{31}=(\mathbf{p}_{31}\cdot\mathbf{E}_s)/(2\hbar)$ with $\mathbf{p}_{21(31)}$ being the dipole transition operator for the signal field $\mathbf{E}_s$. Another weak probe field $E_p$ with angular frequency $\omega_p$ then drives the transition $|2\rangle\leftrightarrow|4\rangle$ and $|3\rangle\leftrightarrow|4\rangle$ simultaneously. Correspondingly, the half-Rabi frequencies of the probe field are defined as $\Omega_{24}=(\mathbf{p}_{24}\cdot\mathbf{E}_p)/(2\hbar)$ and $\Omega_{34}=(\mathbf{p}_{34}\cdot\mathbf{E}_p)/(2\hbar)$ with $\mathbf{p}_{24(34)}$ being the dipole transition operator for the probe field $\mathbf{E}_p$. Detunings are defined by $\Delta_2=\omega_s-(E_2-E_1)/\hbar$, $\Delta_3=\omega_s-(E_3-E_1)/\hbar$ and $\Delta_4=\omega_s-\omega_p-(E_4-E_1)/\hbar$ respectively. Here, $E_j\ (j=1-4)$ is the eigen energy of state $|j\rangle$. Defining the energy splitting $\Delta=(E_3-E_2)/\hbar$, we can express the detunings as $\Delta_2=\Delta/2+\delta$, $\Delta_3=\delta-\Delta/2$ with $\delta=\omega_s-(E_2+E_3-2E_1)/(2\hbar)$.

Under the rotating-wave approximation (RWA), the equations of motion for density-matrix operator $\sigma_{ij}$ in the interaction picture are given by
\begin{subequations}\label{eq:sigma}
\begin{eqnarray}
i\dot{\sigma}_{21}&=&-d_{21}\sigma_{21}-\Omega_{21}(\sigma_{11}-\sigma_{22})-i\kappa\sigma_{31}-\Omega_{24}\sigma_{41} \nonumber\\
& &+\Omega_{31}\sigma_{23}\\
i\dot{\sigma}_{31}&=&-d_{31}\sigma_{31}-\Omega_{31}(\sigma_{11}-\sigma_{33})-i\kappa\sigma_{21}-\Omega_{34}\sigma_{41} \nonumber\\
& &+\Omega_{21}\sigma_{32}\\
i\dot{\sigma}_{41}&=&-d_{41}\sigma_{41}-\Omega_{24}^\ast\sigma_{21}-\Omega_{34}^\ast\sigma_{31}+\Omega_{21}\sigma_{42} \nonumber\\
& &+\Omega_{31}\sigma_{43}\\
i\dot{\sigma}_{32}&=&-d_{32}\sigma_{32}-i\kappa(\sigma_{22}-\sigma_{33})-\Omega_{31}\sigma_{12}-\Omega_{34}\sigma_{42} \nonumber\\
& &+\Omega_{21}^\ast\sigma_{31}+\Omega_{24}^\ast\sigma_{34}\\
i\dot{\sigma}_{42}&=&-d_{42}\sigma_{42}-\Omega_{24}^\ast(\sigma_{22}-\sigma_{44})-\Omega_{34}^\ast\sigma_{32}+\Omega_{21}^\ast\sigma_{41} \nonumber\\
& &+i\kappa\sigma_{43}\\
i\dot{\sigma}_{43}&=&-d_{43}\sigma_{43}-\Omega_{34}^\ast(\sigma_{33}-\sigma_{44})-\Omega_{24}^\ast\sigma_{23}+\Omega_{31}^\ast\sigma_{41} \nonumber\\
& &+i\kappa\sigma_{42}
\end{eqnarray}
\end{subequations}
where dot above $\sigma_{ij}$ denotes the time derivation, $d_{ij}=\Delta_i-\Delta_j+i\gamma_{ij}$ with $\Delta_j$ being the detuning of state $|j\rangle$ and $\gamma_{ij}=(\Gamma_i+\Gamma_j)/2+\gamma_{ij}^{\rm dph}$ respectively. $\Gamma_{j}$ is the total population decay rate of state $|j\rangle$, and $\Gamma_{ij}^{\rm dph}$ is the dephasing rate of coherence $\sigma_{ij}$, which may originate not only from electron-electron scattering and electron-photon scattering, but also from inhomogeneous broadening due to scattering on interface roughness.
$\kappa=\eta\sqrt{\Gamma_2\Gamma_3}$ is the tunnelling interference, which represents the cross coupling of states $|2\rangle$ and $|3\rangle$ contributed by the process in which a phonon is emitted from state $|2\rangle$ and recaptured by state $|3\rangle$. Here, parameter $\eta$ represents the coupling strength of the tunnelling interference~\cite{Yang2014,Liu2000,Faist1997,Schmidt1997}.
%
%

In general, Eqs.~(\ref{eq:sigma}) can be solved in the weak-filed limit. To this end, we assume that the population is initially occupied in the ground state $|1\rangle$, and the state $|1\rangle$ will not be depleted during the time evolution because the signal field is weak and off-resonant in our scheme. In this situation  Eqs.~(\ref{eq:sigma}) can be solved perturbatively by introducing the asymptotic expansion $\sigma_{ij}=\sum^{\infty}_{n=0}\epsilon^{n}\sigma_{ij}^{(n)}$, $\Omega_{j1}=\epsilon\Omega_{j1}^{(1)}$ and $\Omega_{j4}=\epsilon\Omega_{j4}^{(1)}$ with $\sigma_{11}^{(0)}=1$ and $\sigma_{jj}^{(0)}=0\ (j\neq1)$. Here, $\epsilon$ is a small parameter characterizing the small depletion of the ground state. Substituting these expansions into Eqs.~(\ref{eq:sigma}) one obtains a set of linear but inhomogeneous equations of $\sigma_{ij}^{(n)}$, which can be solved order by order.

Using the standard differential Fourier transform technique as shown in Ref. \cite{Zhu2009} one can easily obtain the leading order ($n=1$) solution, which are given by
\begin{subequations}\label{eq:1st}
\begin{eqnarray}
& & R_{21}^{(1)}=\frac{-\Lambda_{21}^{(1)}(\omega+d_{31})+i\kappa\Lambda_{31}^{(1)}}{(\omega+d_{21})(\omega+d_{31})+\kappa^2},\\
& &R_{31}^{(1)}=\frac{-\Lambda_{31}^{(1)}(\omega+d_{21})+i\kappa\Lambda_{21}^{(1)}}{(\omega+d_{21})(\omega+d_{31})+\kappa^2},
\end{eqnarray}
\end{subequations}
and $R_{41}^{(1)}=R_{32}^{(1)}=R_{42}^{(1)}=R_{43}^{(1)}=0$. Here, $R_{ij}^{(n)}$ and $\Lambda_{ij}^{(1)}$ are the Fourier transforms of $\sigma_{ij}^{(n)}$ and $\Omega_{ij}^{(1)}$, respectively. $\omega$ is Fourier transformation variable.

Taking $\Lambda_{21(31)}^{(1)}=p_{21(31)}\Lambda^{(1)}_s/(2\hbar)$ with $\Lambda^{(1)}_s$ being the Fourier transform of $E_s$, one can express the dispersion relation for the signal field as
\begin{eqnarray}
W_s(\omega)=\frac{\omega}{c}+\frac{{\cal N}_e\omega_s}{2\varepsilon_0c}\times\frac{p_{12}R_{21}^{(1)}+p_{13}R_{31}^{(1)}}{\Lambda^{(1)}_s},
\end{eqnarray}
where $c$ is the light speed in vacuum and $\varepsilon_0$ is the vacuum electrical conductivity.

In most cases, $W_s(\omega)$ can be expanded in a McLaurin series around the center frequency of the probe field (i.e., $\omega=0$)
\begin{equation}
W_s(\omega)=W_s^{(0)}+W_s^{(1)}\omega+\frac{1}{2}W_s^{(2)}\omega^2+O(\omega^3),
\end{equation}
where $W_s^{(j)}=(\partial^j W_s/\partial \omega^j)|_{\omega=0}$. Here, $W_s^{(0)}=\phi+i\alpha/2$ describes the phase shift $\phi$ per unit length and the linear absorption coefficient $\alpha$ of the signal field. $W_s^{(1)}=1/V_g$ gives the propagation velocity, and $W_s^{(2)}$ represents the group velocity dispersion which contributes both to the pulse shape change and additional loss of signal field intensity.

\begin{figure}[htb]
  \centering
  \includegraphics[width=8.5cm]{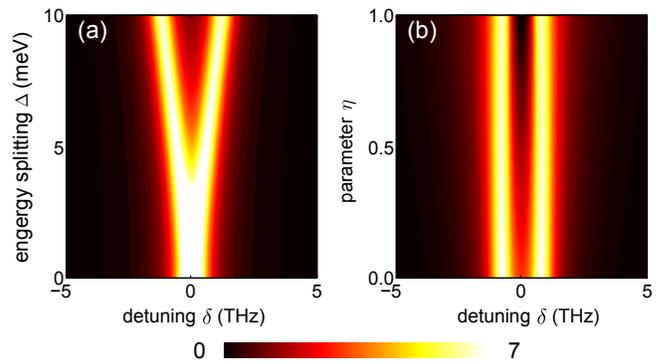}\\
  \caption{(Color online) Contour plots of the absorption coefficient $\alpha$ of the signal field: Panel (a), as a function of detuning $\delta$ and the energy splitting $\Delta$; Panel (b), as a function of detuning $\delta$ and the parameter $\eta$. Bright areas correspond to low absorption, dark to large absorption. Other system parameters are given in the text.}\label{fig:Ws}
\end{figure}
Figure \ref{fig:Ws} shows contour plots of the absorption coefficient $\alpha$ of the signal field as functions of the detuning $\delta$ and  the energy splitting $\Delta$ (see panel (a)), as well as the detuning $\delta$ and parameter $\eta$ (see panel (b)). Here, $\eta=0$ denotes no tunnelling interference, and $\eta=1$ corresponds to perfect tunnelling interference. In panel (a), we take  $\Delta_4=0$, $\eta=0$,  $\Gamma_2=\Gamma_3=0.68$ meV and $\Gamma_4=0.01\Gamma_2$ and $\gamma_{ij}^{\rm dph}=1.2$ meV except for $\gamma_{14}^{\rm dph}\approx0$ meV because of the high inter-well barrier between states $|1\rangle$ and $|4\rangle$~\cite{Zhu2011} ($1$ meV is equivalent to $0.24$ THz). The electric-dipole moment is $|p_{j1}|=|p_{j4}|=8.5\times10^{-28}$ C$\cdot$m ($j=2,3$). It is evident that for small energy splitting between states $|2\rangle$ and $|3\rangle$ the absorption profile has only a single peak, which result in a significant attenuation of the signal field at $\delta=0$. However, when increasing the energy splitting, we can see that the absorption profile exhibits a large EIT-like absorption doublet, which is known as the tunnelling induced transparency (TIT). In panel (b)	we take $\Delta=6.5$ meV and choose the parameter $\eta$  as a variable. Clearly, the absorption of the signal field at $\delta=0$ is significantly reduced with the increase of the parameter $\eta$. To show the physical mechanism of the reduction of the signal field absorption, we use the spectrum-decomposition method~\cite{Zhu2013} to analyze the characteristics
of the signal-field absorption $\alpha$ explicitly.

Assuming $p_{21}=p_{31}=p_0$ for simplicity, one can obtain
\begin{eqnarray}\label{eq:alpha}
\alpha&=&{\rm Im}\left\{\kappa\frac{i\kappa-\left[\delta+i(\gamma_{21}+\gamma_{31})/2\right]}{(\delta+\Delta/2+i\gamma_{21})(\delta-\Delta/2+i\gamma_{31})+\kappa^2}\right\}\nonumber\\
&=& \frac{\kappa}{2}\left\{\left[\frac{\Gamma}{(\delta+\delta_0)^2+\Gamma^2}+\frac{\Gamma}{(\delta-\delta_0)^2+\Gamma^2}\right]\right.\nonumber\\
& & + \left.\frac{g}{\delta_0}\left[\frac{\delta-\delta_0}{(\delta-\delta_0)^2+\Gamma^2}-\frac{\delta+\delta_0}{(\delta+\delta_0)^2+\Gamma^2}\right]\right\}.
\end{eqnarray}
where $\kappa={\cal N}_e\omega_s|p_0|^2/(c\varepsilon_0\hbar)$, $\delta_0=\sqrt{(\Delta/2)^2-\Gamma^2}$,  $\Gamma=(\gamma_{21}+\gamma_{21})/2$ and $g=\kappa$. Obviously, terms in the first square bracket on the right hand side of Eq. (\ref{eq:alpha}) are two Lorentzians, which are the net contribution to signal-field absorption from two different channels corresponding to the two excited states with  $\Gamma$ being the width (also strength) of the two Lorentzians and  $\delta_0$ being the real part of the spectrum poles. The following terms in the second square bracket are clearly quantum interference ones. It is apparent to see that the magnitude of the interference is controlled by the parameter $g$. If $g>0$ ($g<0$),
the interference is destructive (constructive). In our system, the reduction of the signal field absorption at $\delta=0$ is originated from the destructive interference if $\kappa>0$. Only in the case of $\kappa=0$, the transparency window in the absorption profile is caused by Autler-Townes splitting.

Now, we study the dynamical evolution of the probe field. Firstly, we find the second-order ($n=2$) solution, which are given by
\begin{subequations}\label{eq:2nd}
\begin{eqnarray}
& & R_{41}^{(2)}=\frac{-\Lambda_{24}^{(1)\ast}R_{21}^{(1)}-\Lambda_{34}^{(1)\ast}R_{31}^{(1)}}{\omega+d_{41}},\\
& & R_{32}^{(2)}=\frac{-\Lambda_{31}^{(1)}R_{12}^{(1)}+\Lambda_{21}^{(1)\ast}R_{31}^{(1)}}{\omega+d_{32}},
\end{eqnarray}
\end{subequations}
and $R_{21}^{(2)}=R_{31}^{(2)}=R_{24}^{(2)}=R_{34}^{(2)}=0$.

At the third-order ($n=3$), we can easily obatin the solutions for $\sigma_{42}$ and $\sigma_{43}$ respectively, which reads
\begin{subequations}\label{eq:3rd}
\begin{eqnarray}
R_{24}^{(3)}&=&\frac{1}{D}\left[(\omega+d_{34})(\Lambda_{34}^{(1)}R_{23}^{(2)}-\Lambda_{21}^{(1)}R_{14}^{(2)})\right.\nonumber\\
& &\left.-i\kappa(\Lambda_{24}^{(1)}R_{32}^{(2)}-\Lambda_{31}^{(1)}R_{14}^{(2)})\right],\\
R_{34}^{(3)}&=&\frac{1}{D}\left[(\omega+d_{24})(\Lambda_{24}^{(1)}R_{32}^{(2)}-\Lambda_{31}^{(1)}R_{14}^{(2)})\right.\nonumber\\
& &\left.-i\kappa(\Lambda_{34}^{(1)}R_{23}^{(2)}-\Lambda_{21}^{(1)}R_{14}^{(2)})\right],
\end{eqnarray}
\end{subequations}
where $D=(\omega+d_{24})(\omega+d_{34})+\kappa^2$ and $R_{ij}^{(n)}=R_{ji}^{(n)\ast}$.

The dynamical evolution of the probe field is governed by the Maxwell equation, i.e.,
\begin{eqnarray}
& & \nabla^2\mathbf{E}_p-\frac{1}{c^2}\frac{\partial^2 \mathbf{E}_p}{\partial t^2}=\frac{1}{\varepsilon_0c^2}\frac{\partial^2\mathbf{P}}{\partial t^2},
\end{eqnarray}
where $\mathbf{E}_p=\mathbf{e}_pE_p\exp{[i(k_pz-\omega_pt)]}$ with $\mathbf{P}=\mathbf{e}_p{\cal N}_e(p_{24}\sigma_{42}+p_{34}\sigma_{43})\exp{[i(k_pz-\omega_pt)]}$. Here, $\mathbf{e}_{p}$ is the polarization vector of the electric field. Under the slow-varying envelope approximation (SVEA), the Maxwell equation is reduced to
\begin{align}
&i\left(\frac{\partial}{\partial z}+\frac{1}{c}\frac{\partial }{\partial t}\right)E_p+\frac{{\cal N}_e\omega_p}{2\varepsilon_0c}\left(p_{42}\sigma_{24}+p_{43}\sigma_{34}\right)=0.
\end{align}
Taking the time-Fourier transformation and inserting the solutions of Eqs. (\ref{eq:sigma}), we obatin
\begin{align}\label{eq:Ep}
& \frac{\partial}{\partial z}\Lambda_p^{(1)}-i\frac{\omega}{c}\Lambda_p^{(1)}=i\frac{{\cal N}_e\omega_p}{2\varepsilon_0c}\left(p_{24}R_{42}^{(3)}+p_{34}R_{43}^{(3)}\right).
\end{align}
where $\Lambda^{(1)}_p$ is the Fourier transform of $E_p$. The right-hand side of the above equation is known as the nonlinear term (NLT), which can be expressed as
\begin{eqnarray}
{\rm Nonlinear\ term}&=&\frac{{\cal N}_e\omega_p}{2\varepsilon_0c}\left(p_{24}R_{42}^{(3)}+p_{34}R_{43}^{(3)}\right)\nonumber\\
&=&(\phi_{\rm XPM}+i\alpha_{\rm NL})\Lambda_p^{(1)},
\end{eqnarray}
where we have defined the nonlinear phase shift $\phi_{\rm XPM}$ (i.e., Kerr cross-phase modulation) per unit length and the third-order nonlinear absorption coefficient $\alpha_{\rm NL}$ respectively.

%
\begin{figure}[htb]
  \centering
  \includegraphics[width=8.5cm]{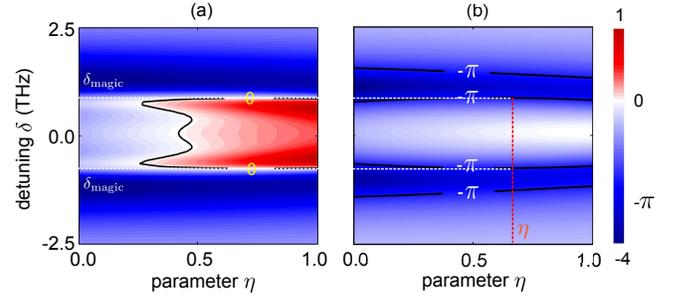}\\
  \caption{(Color online) Contour plots of the third-order gain/loss $\alpha_{\rm NL}$ (panel (a)) and the cross-phase modulation $\phi_{\rm XPM}$ (panel (b)) of the probe field as functions of the detuning $\delta$ and the parameter $\eta$. The equal-altitude lines (zero for $\alpha_{\rm NL}$ and $-\pi$ for $\phi_{\rm XPM}$) guide the selection of a magic detuning $\delta_{\rm magic}$ (the horizontal
dashed line). The vertical line indicates that $\eta\approx0.7$ is required to achieve the Kerr phase gate operation.}\label{fig:Wp}
\end{figure}
In Figure \ref{fig:Wp},
we plot the Kerr phase $\phi_{\rm XPM}$ and the third-order gain/loss $\alpha_{\rm NL}$ as functions of the detuning $\delta$ and the parameter $\eta$ respectively. The system parameters are the same as those used in Fig. \ref{fig:Ws}. In Fig. \ref{fig:Wp}(a), we show that in the absence of the tunnelling interference (i.e., $\eta=0$), the coefficient $\alpha_{\rm NL}$ is negative which corresponds to a large amplification for the probe field. However, if increasing the tunnelling interference, one can see that the third-order gain decreases dramatically and changes to be positive which results in a third-order absorption for the probe field. Therefore, there exist a specific signal field detuning defined as a 'magic' detuning $\delta_{\rm magic}$, at which the probe field will not be amplified/absorbed (i.e., $\alpha_{\rm NL}=0$). Simultaneously, $\phi_{\rm XPM}=-\pi$ is achieved with a specific parameter $\eta$ (see panel (b)), then at the exit of the medium, the phase of the probe field will have undergone a perfect 180$^\circ$ rotation.

\begin{figure}[htb]
  \centering
  \includegraphics[width=8.5cm]{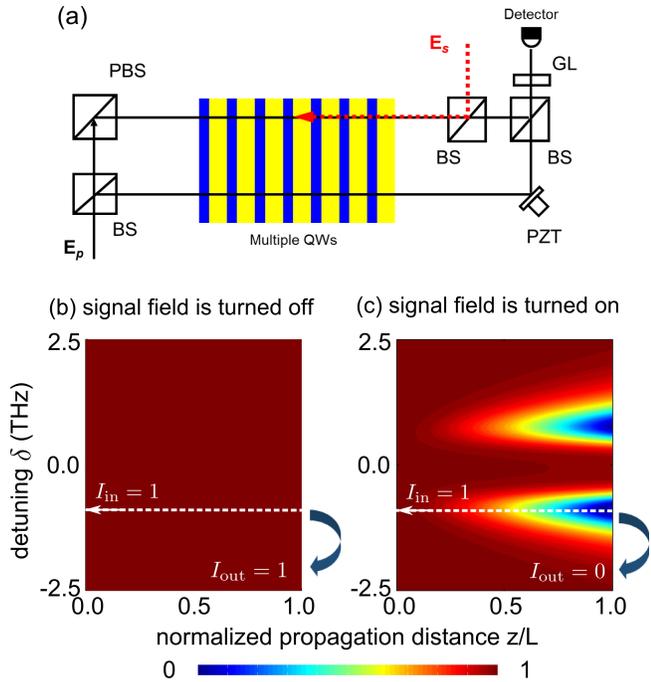}\\
  \caption{(Color online) The layout of the experimental setup (panel (a)) and plots of $I_p(z,\delta)$ as functions of $\delta$ and the normalized propagation distance $z/L$. The dashed line indicates a magic signal-field detuning with which a lossless Kerr phase gate operation can be achieved. Left plot: the signal field is turned off, $I_{\rm in}=1$ and $I_{\rm out}=1$ at the exit of the medium $z/L=1$ (panel (b)). Right plot: the signal field is turned on, $I_{\rm in}=1$ and $I_{\rm out}=0$ at the exit of the medium $z/L=1$ (panel (c)). The Mach-Zehnder interferometers, controlled by PZT, are used to observe the nonlinear phase shift for the probe field.}\label{fig:Ep}
\end{figure}
To verify the above analysis, we suggest an experimental scheme (see Fig.~\ref{fig:Ep}(a)) to realize a zero to $-\pi$ lossless Kerr phase gate operation by using a Mach-Zehnder interferometer. As shown in Fig.~\ref{fig:Ep}(a), the probe field is split by a beam splitter (BS), and one is probe field while another is reference light. They are combined together using another BS to build the Mach-Zehnder interferometer. The signal light is overlapped with the probe field with the opposite direction (see Fig.~\ref{fig:Ep}(a)). Before the detector, a Glan-Taylor prism (GL) is used to filter the signal field. By performing full numerical simulations using equation (\ref{eq:Ep}), we show that a zero to $-\pi$ Kerr phase gate operation can be achieved in our system. Figure \ref{fig:Ep}(b) and (c) display two contour plots that show the intensity of the probe field as functions of the detuning $\delta$ and the normalized propagation distance $z/L$ in the medium when the signal light is turn on/off. The white dashed line indicates a ¡°magic¡± detuning $\delta_{\rm magic}$ at which the third-order gain/loss is eliminated due to the tunnelling interference effect, and simultaneously a $-\pi$ cross phase modulation is achieved. Both contour plots are normalized with respect to the probe field amplitude at the entrance of the medium, i.e., $I_{\rm in}(z=0)=1$. Correspondingly, we have $I_{\rm out}=1$ (the probe field and reference field have the same phase shift) at the exit $z/L=1$ when the signal light is turned off (see panel (b)). However, if the signal field is added, we have $I_{\rm out}=0$ (see panel (c)). Therefore, a zero to $-\pi$ lossless Kerr phase gate operation is achieved.

In conclusion, we have demonstrated theoretically a weak-light, zero to $\pi$ controllable nonlinear Kerr phase gate in semiconductor quantum wells structure. We show that a Autler-Townes-like splitting in the linear absorption spectrum appears due to the tunnelling interference. Using the spectrum-decomposition method, we show that the tunnelling effect attributes to a destructive interference, which results in a deeper tunneling induced transparency window. We further show that it is possible to find a magic detuning for a signal field so that the probe field acquires a $-\pi$ phase rotation due to the tunnelling interference in the QWs system. Our numerical calculations have shown that the schemes and methods studied can indeed lead to a zero to $-\pi$ phase-gate
operation, which may find many applications in optical telecommunications.

We thank Dr. L. Deng for useful discussion and suggestions.

\end{document}